\documentclass{desyproc}

\begin{document}
\title{Axion Haloscopes with Toroidal Geometry \\
  at CAPP/IBS}

\author{{\slshape  Byeong Rok Ko}\\[1ex]
  Center for Axion and Precision Physics Research, Institute for Basic
  Science (IBS), \\
  Daejeon 34141, Republic of Korea}

\contribID{familyname\_firstname}

\confID{13889}  
\desyproc{DESY-PROC-2016-XX}
\acronym{Patras 2016} 
\doi  

\maketitle

\begin{abstract}
  \hspace{3ex}
  The present state of the art axion haloscope employs a cylindrical
  resonant cavity in a solenoidal field. We, the Center for Axion and
  Precision Physics Research (CAPP) of the Institute for Basic Science
  (IBS) in Korea, are also pursuing halo axion discovery using this
  cylindrical geometry. However, the presence of end caps of cavities
  increases challenges as we explore higher frequency regions for the
  axion at above 2 GHz. To overcome these challenges we exploit a
  toroidal design of cavity and magnetic field. A toroidal geometry
  offers several advantages, two of which are a larger volume for a
  given space and greatly reduced fringe fields which interfere with
  our  preamps, in particular the planned quantum-based devices. We
  introduce the concept of toroidal axion haloscopes and present
  ongoing  research activities and plans at CAPP/IBS.  
\end{abstract}

\section{Axion haloscopes at CAPP/IBS}
\hspace{3ex}
One of the primary targets of the Center for Axion and Precision
Physics research (CAPP) of the Institute for Basic Science (IBS) in
Korea is to search for axion cold dark matter using the method
proposed by Sikivie~\cite{sikivie}, also known as the axion haloscope
search. We, CAPP/IBS, are pursuing traditional axion haloscopes with
cylindrical geometry, where two significant solenoids are
employed. One has 12 T of $B$ field and 32 cm of cold bore and the
other has 25 T of $B$ field and 10 cm of cold bore. The former can
scan the axion frequency from 0.5 to 1.3 GHz, while the latter can
scan from 1.8 to 10 GHz. The details of halo axion searches using
these cylindrical geometry will be given in the near future.

In the same pipe line, we are also considering axion haloscopes with
toroidal geometry which offer a larger volume and greatly reduced
fringe $B$ fields. In view of these advantages, we introduce currently
ongoing research activities and plans for toroidal axion haloscopes at
CAPP/IBS in this proceedings.

\section{Cylindrical geometry vs. toroidal geometry}
\hspace{3ex}
We compare two axion haloscopes with two different geometry,
cylindrical and toroidal, to reveal the advantages of toroidal axion
haloscopes mentioned above.
Cylindrical cavities have an electromagnetic cavity mode parallel to
an external static magnetic field $\vec{B}_{\rm external}$ provided by
the magnet employed in an axion haloscope, which is the TM$_{010}$
mode, and so do toroidal cavities, which is to be referred to as
Quasi-TM (QTM) mode~\cite{QTM} in this proceedings.
As reported in Ref.~\cite{BRKo:2016},
their electric ($C_E$) and magnetic ($C_M$) form factors are also the
same regardless of the cavity location inside the magnet.

\begin{wrapfigure}{r}{0.6\textwidth}
  \centerline{\includegraphics[width=0.55\textwidth]{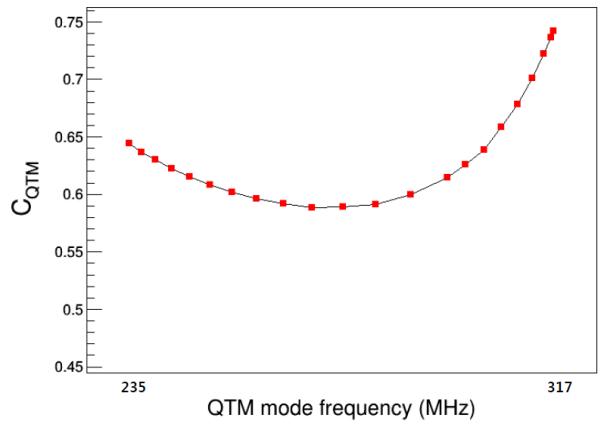}}
  \caption{Form factors of the QTM mode of the toroidal cavity as a
    function of frequency.}  
  \label{Fig:modes}
\end{wrapfigure}

In the absence of end caps of cavities in a toroidal design, the
cavity modes depend only on the minor radius of the toroidal
cavity. Therefore, degeneration of the cavity modes is expected to be
little and indeed the finite element method~\cite{CST} demonstrates no
degeneracy, or equivalently no mode crossing, in the QTM mode as shown
in Fig.~\ref{Fig:modes}.

\begin{table}
\centerline{\begin{tabular}{|c|c|c|}\hline
characteristics           & cylindrical geometry & toroidal geometry\\ \hline\hline
$\vec{B}_{\rm external}$  & $\sim B_0\hat{z}$& $\sim\frac{B_0}{\rho}\hat{\phi}$ \\\hline
                 &independent of     &depends on  \\
$B^2_{\rm avg}V$ &the cavity location&the cavity location \\
                 &inside a solenoid  &inside a toroidal magnet \\\hline
cavity mode $\Arrowvert~\vec{B}_{\rm external}$  & TM$_{010}$& QTM \\\hline
form factor  &$C_E=C_M\propto B^2_{\rm avg}V$ &$C_E=C_M\propto B^2_{\rm avg}V$ \\\hline
                &                &avoidable   \\
$B_{\rm fringe}$&unavoidable     &with  \\
                &                &additional coils  \\\hline
mode crossing  &Yes&No\\\hline
\end{tabular}}
\caption{Comparison of characteristics between cylindrical and
  toroidal axion haloscopes, where $B_0$ is a constant, $B^2_{\rm
    avg}V\equiv\int\vec{B}^2_{\rm external}dV$, and $V$ is the cavity
  volume.}
\label{tab:comp}
\end{table}
\noindent This {\it no mode crossing} in toroidal
geometry enables us to increase the cavity volume very effectively.

The fringe $B$ fields, from toroidal magnets, $B_{\rm fringe}$ are
ideally zero and even practically very small compared to those from
solenoids. Furthermore, the $B_{\rm fringe}$ from toroidal magnets can
be reduced with additional coils. With greatly reduced $B_{\rm
  fringe}$, the handling of quantum preamps in toroidal geometry is
much easier than that in cylindrical geometry.

Table~\ref{tab:comp} summarizes the comparison between axion
haloscopes with cylindrical and toroidal geometry. Note that $z$,
$\rho$, and $\phi$ refer to cylindrical coordinates.

\section{Axion haloscopes with toroidal geometry at CAPP/IBS}
\begin{wrapfigure}{r}{0.6\textwidth}
  \centerline{\includegraphics[width=0.5\textwidth]{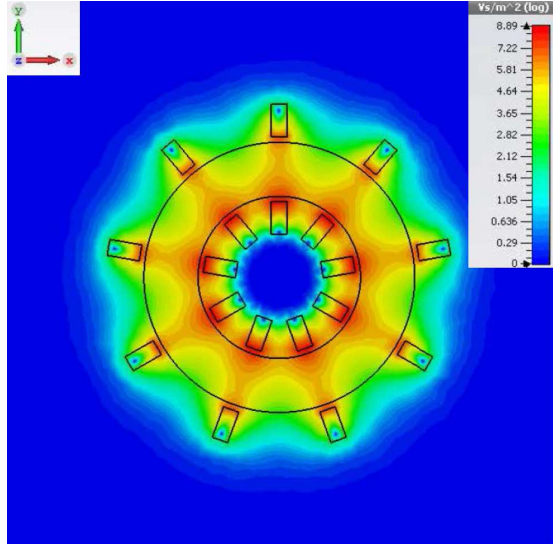}}
  \caption{Conceptual design of the large toroidal magnet with
  9 coils and the cavity. The color map shows the $B$ field from the
  magnet.}
  \label{Fig:LargeToroid}
\end{wrapfigure}
\hspace{3ex}
We are considering the two toroidal axion haloscopes whose magnet
specifications are shown in Table~\ref{tab:toroid}.

\begin{figure}[htbp]
  \centerline{\includegraphics[width=0.8\textwidth]{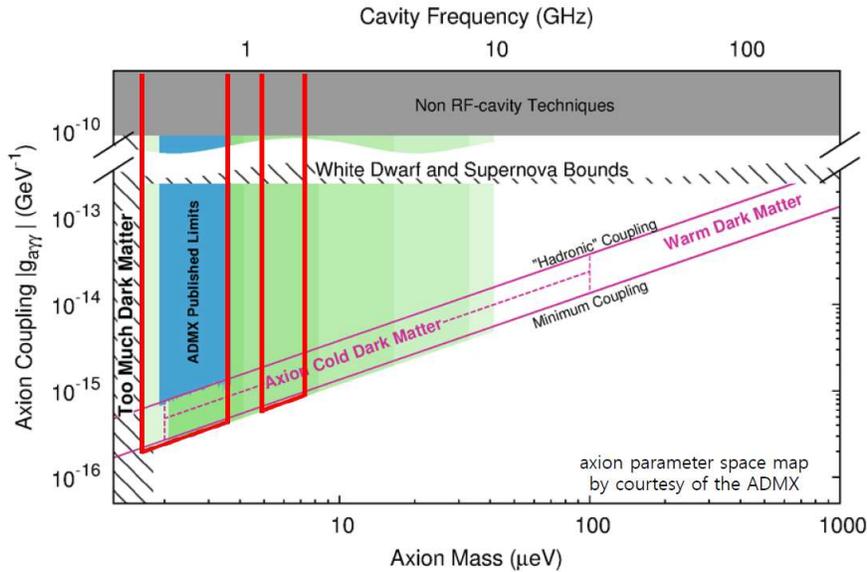}}
  \caption{Axion parameter space where red lines in the lower mass
    region denote the search region of the large toroidal axion
    haloscopes and those in the higher mass region denote that of the
    small toroidal axion haloscopes at CAPP/IBS.}  
  \label{Fig:Sensitivity}
\end{figure}

The large toroidal magnet illustrated in Fig.~\ref{Fig:LargeToroid} can hold
a cavity with 200 cm of major radius and 50 cm of minor radius, which
enables us to scan the axion parameter space down to 190 MHz with a
dielectric ($\epsilon_r = 9.9$) tuning rod. Thanks to 9,870 liters of
the cavity volume which is about 50 times larger than the ADMX
cavity~\cite{ADMX}, we can achieve a reasonable scanning
rate~\cite{SCANNING_RATE} even with a semiconductor-based preamp and
cavity cooling with liquid helium (LHe) which results in the
relatively high system noise of $\sim$6 K, with 2 K from the preamp
and 4 K from the cavity. The large toroidal magnet also
can hold 4 cavities, each of them having 200 cm of major radius and 20 cm
of minor radius, which enables us to scan the axion parameter space
up to 850 MHz with a conductor tuning rod. With the 4-cavity system,
we can achieve a reasonable scanning rate even with a system noise of
$\sim$6 K. Figure~\ref{Fig:Sensitivity} shows the relevant
parameter space and expected sensitivity from the large toroidal axion
haloscopes which will be realized in about ten years.
\begin{wraptable}{l}{0.5\textwidth}
\centerline{\begin{tabular}{|c|c|c|}\hline
                 & small toroid & large toroid\\ \hline\hline
$B_{\rm avg}$    & 12 T  & 5 T \\\hline
number of coils  & 36    & 9 \\\hline
major radius     & 50 cm & 200 cm \\\hline
minor radius     & 11 cm & 60 cm \\
\hline
\end{tabular}}
\caption{Specifications of the two toroidal magnets.}
\label{tab:toroid}
\end{wraptable}

The small toroidal magnet can hold a cavity with 50 cm of major radius
and 9 cm of minor radius, which enables us to search for axion
parameter space from 1.3 to 1.8 GHz with a conductor tuning rod. The
cavity can be cool down to 100 mK with a dilution fridge. Then, we
can be sensitive to KSVZ~\cite{KSVZ1,KSVZ2} axion search using a semiconductor
preamp or DFSZ~\cite{DFSZ1,DFSZ2} using a quantum version preamp with
acceptable scanning rates. Figure~\ref{Fig:Sensitivity} also shows the
relevant parameter space and expected sensitivity from the small
toroidal axion haloscopes which will be realized in a few years.

\section{Summary}
\hspace{3ex}
CAPP/IBS realizes several advantages in axion haloscopes with toroidal
geometry, two of which are a larger volume for a given space and
greatly reduced fringe fields. Thanks to the very large volume of the
cavity in the large toroidal axion haloscopes, we can realize a
reasonable scanning rate even with 4 K cavity cooling which can be
achieved easily with LHe. The capability of cooling a system using LHe
shows how powerful the large toroidal axion haloscopes are. In the
near future, we will realize the small toroidal axion haloscopes to
demonstrate the feasibility of the large toroidal axion haloscopes.
\section*{Acknowledgments}
\hspace{3ex}
This work was supported by IBS-R017-D1-2016-a00. 
\begin{footnotesize}

\end{footnotesize}



\begin{thebibliography}{99}
\bibitem{QTM}
  Not the same as TM$_{010}$ mode, but similar to it.
\end{thebibliography}
\end{document}